\documentclass[11pt]{article}
\usepackage{verbatim}
\usepackage{fullpage}
\usepackage{amsfonts}
\usepackage{amssymb}
\usepackage{color}
\usepackage{graphicx}
\usepackage{helvet}
\usepackage{latexsym}
\usepackage{amsgen}
\usepackage{amsmath}
\usepackage{amsopn}
\usepackage{amsxtra}

\newtheorem{theorem}{Theorem}

\newtheorem{proposition}[theorem]{Remark}

\newcommand{\ket}[1]{|#1\rangle}
\newcommand{\braket}[2]{\langle #1|#2\rangle}

\renewcommand{\H}{\mathcal{H}}

\begin{document}

\title{Two remarks on the local Hamiltonian problem\thanks{This
  material is based upon work supported by NSF Grant \#0523866
  while the author was at Rutgers University.}}
\author{Peter C. Richter\thanks{Laboratoire de Recherche en
  Informatique, Universit\'{e} de Paris-Sud XI, Orsay, France.  {\tt
  richterp@lri.fr}}}
\date{}
\maketitle

\begin{abstract}
In this note we present two natural restrictions of the local
Hamiltonian problem which are BQP-complete under Karp
reduction.  Restrictions complete for QCMA, QMA$_1$, and MA were
demonstrated previously.
\end{abstract}


\paragraph{Introduction.}
The complexity class BQP captures those problems solvable in
polynomial time by bounded-error quantum
algorithms.  BQP has complete ``promise problems'' -- for example,
determining the sign of a quadratically signed weight enumerator
\cite{KL}, obtaining additive approximations to the Jones polynomial
\cite{AJL} and the Tutte polynomial \cite{AAEL} evaluated at specific
points, estimating the diagonal entries of a matrix power \cite{JW},
and sampling from the energy eigenvalues of a local Hamiltonian
\cite{WZ}.

In this note we demonstrate two promise problems which are (a)
BQP-complete under {\em Karp reduction}, or polynomial-time
transformation, and (b) natural restrictions of the canonical
QMA-complete ``local Hamiltonian problem'' \cite{KSV} likely to arise
in computational physics applications.  Our observations are in the
spirit of results \cite{WJB,Bra,BBT} which demonstrated restrictions
of the local Hamiltonian problem complete for QCMA, QMA$_1$ and MA.

\paragraph{Definitions.}
A {\em promise problem} $P$ is the disjoint union of two sets $L_0,
L_1 \subseteq \{0,1\}^*$.  A bounded-error algorithm $A$ {\em
decides} $P$ if its output bit $A(x)$ satisfies:
\begin{equation}
x \in L_b \Rightarrow \Pr[A(x) = b] \geq 2/3
\end{equation}
Inputs $x \notin L_0 \cup L_1$ are ``promised'' not to occur;
equivalently, they can cause $A$ to behave arbitrarily.  If $A$ is a
quantum (resp., classical) algorithm running in time $O(poly(|x|))$,
then $P$ is in BQP (resp., BPP).  A BQP {\em verifier} $V$ takes both
an input $x \in L_0 \cup L_1$ and a quantum {\em witness} state
$\ket{\psi}$ on $O(poly(|x|))$ qubits and outputs a bit
$V(x,\ket{\psi})$ satisfying:
\begin{eqnarray}
x \in L_1 & \Rightarrow & \exists \ket{\psi} \: : \:
\Pr[V(x,\ket{\psi}) = 1] \geq 2/3 \label{c-param}\\
x \in L_0 & \Rightarrow & \forall \ket{\psi} \: : \:
\Pr[V(x,\ket{\psi}) = 1] \leq 1/3 \label{s-param}
\end{eqnarray}
If there is a BQP verifier for $P$, then $P$ is in QMA.  We obtain the
complexity class (a) QMA$_1$ by changing the completeness parameter
in line (\ref{c-param}) from $2/3$ to $1$, (b) QCMA by restricting the
(witness, verifier) pair to be (classical, quantum), (c) MA by
restricting the pair to be (classical, classical), and (d) NP by
restricting the pair to be (classical, classical) and changing the
soundness parameter in line (\ref{s-param}) from $1/3$ to $0$.

Let $G=(V,E)$ be a hypergraph whose $n$ vertices are $d$-state spins
(``qudits'' with $d$-dimensional Hilbert space $\H_d$) and whose
hyperedges are $k$-subsets of qudits.  We shall assume that $d$ and
$k$ are fixed independently of the scaling parameter $n$.  Denote by
$H_e : \H_d^{\otimes k} \rightarrow \H_d^{\otimes k}$ a Hamiltonian
(Hermitian operator) acting on the qudits $e \in E$, and let $I$ be
the identity operator.  A {\em $k$-local Hamiltonian} has the form:
\begin{equation}
H = \sum_{e \in E} H_e \otimes I_{V \setminus e}
\end{equation}
Its {\em ground state energy} is its minimum eigenvalue $\lambda_1$,
and an eigenvector with eigenvalue $\lambda_1$ is a {\em ground
state}.  The {\em local Hamiltonian problem} (LH-MIN) is to decide if
$\lambda_1$ is at most $a$ (``YES'') or at least $b = a + \Delta$
(``NO''), where $\Delta = \Omega(1/poly(n))$ is the {\em promise gap}
and $\lambda_1$ is promised not to lie in $(a,b)$.  The inequality
$\lambda_1 \leq a$ can be demonstrated by verifying the existence of a
{\em low-energy state} for $H$ -- i.e., a state $\ket{\psi}$
satisfying $\braket{\psi}{H|\psi} \leq a$.  LH-MIN is QMA-complete by
Kitaev \cite{KSV}.

\paragraph{A low-energy state promise.}
Let $U = U_L \cdots U_2 U_1$ be an $n$-qubit, $L$-gate BQP circuit
with input $x \in L_b$ (hardwired into the first few gates)
that maps the classical bitstring $\ket{00\cdots0}$ to a quantum state
whose ``answer qubit'' outputs $b$ with probability at least $2/3$
when measured.  Using known techniques \cite{KSV,AvKLLR} it is
straightforward to build from $U$ a local Hamiltonian
\begin{equation}\label{kitaev-ham}
H = H_{in} + H_{prop} + H_{clock} + H_{out}
\end{equation}
on $n$ ``circuit'' qubits and $L$ ``clock'' qubits such that (i) the
terms $H_{in}$, $H_{prop}$, and $H_{clock}$ ensure that any low-energy
state of $H$ encodes the circuit computation $U\ket{00\cdots0}$ and
accompanying clock ticks $\ket{1^l 0^{L-l}} \mapsto \ket{1^{l+1}
0^{L-(l+1)}}$ correctly, and (ii) the final term
$H_{out}$ ensures that any low-energy state of $H$ corresponds to an
input $x \in L_1$.  More precisely, there exist parameters $a < b$
with $\Delta = b - a = \Omega(1/poly(n))$ such that
(a) if $x \in L_0$ then every state $\ket{\psi}$ has energy
$\braket{\psi}{H|\psi} \geq b$, and (b) if $x \in L_1$ then the state
\begin{equation}\label{kitaev-state}
\ket{\eta} := \frac{1}{\sqrt{L+1}} \sum_{l=0}^L U_l \cdots U_1
\ket{00\cdots0} \otimes \ket{1^l 0^{L-l}}
\end{equation}
has energy $\braket{\eta}{H|\eta} \leq a$.  Notice that
$|\braket{00\cdots0}{\eta}|^2 \geq \frac{1}{L+1} = \Omega(1/poly(n))$.
Thus, we have demonstrated a Karp reduction from an arbitrary promise
problem in BQP
to a special case of LH-MIN (let us call it LH-MIN$^*$) in which every YES
instance $H$ possesses a low-energy state $\ket{\psi}=\ket{\eta}$ of
``large'' (i.e., size $\Omega(1/poly(n))$) inner product with an {\em
a priori} known classical state -- the bitstring $\ket{00\cdots0}$ in
this case, but we could just as easily have chosen any classical
bitstring.  Furthermore, LH-MIN$^*$ can be solved in BQP using the
Abrams-Lloyd algorithm \cite{AL} (phase estimation on $e^{iH}$) with
$\ket{00\cdots0}$ as the input state.  Thus, we have:
\begin{proposition}
The promise problem LH-MIN$^*$ is BQP-complete under Karp reduction.
\end{proposition}
The promise on YES instances of LH-MIN$^*$ is a natural one and might
be efficiently verifiable for typical inputs using perturbation
theory.  The related problem of {\em sampling} an eigenvalue
$\lambda_k$ of $H$ from the distribution $|\braket{x}{\phi_k}|^2$,
where $\ket{\phi_k}$ is the eigenvector for $\lambda_k$ and $\ket{x}$
is a classical bitstring, is BQP-complete under {\em Cook reduction}
-- i.e., an oracle for the problem can be used by a BPP machine to
solve any problem in BQP \cite{WZ}.

Consider the problem obtained by modifying LH-MIN$^*$ so that for a
YES instance, the classical state of large inner product with a
low-energy state $\ket{\psi}$ is no longer $\ket{00\cdots0}$, but
rather some {\em unknown} classical state $\ket{b_1 b_2 \cdots b_n}$.
Without modification, Kitaev's QMA-completeness theorem for LH-MIN
\cite{KSV} shows that this problem is QCMA-complete.  The problem
remains in QCMA if for a YES instance, $H$ is required only to have a
low-energy state $\ket{\psi}$ of large (size $\Omega(1/poly(n))$)
inner product with some state $\ket{\psi'}$ computable by a
polynomial-size quantum circuit (cf. \cite{WJB}): given
the circuit's description as a witness, the verifier can prepare
$\ket{\psi'}$ and then run Abrams-Lloyd.\footnote{The space of quantum
states is too large to cover with such a fine $\epsilon$-net using
circuits of only polynomial size; otherwise, it would trivially follow
that QCMA=QMA \cite{AK}.}  Similarly, our BQP-complete problem
LH-MIN$^*$ remains so if the state approximating $\ket{\psi}$ is
relaxed from a known classical state to a quantum state having a known
polynomial-time construction.

\paragraph{A spectral gap promise.}
Estimating the ground state energy of a local Hamiltonian $H$ is a
central one in computational physics.  In practice, when the {\em
spectral gap} $\delta := \min_{k \neq 1} \lambda_k - \lambda_1$
of $H$ is large, the problem is often solvable efficiently by a
classical divide-and-conquer ``renormalization group'' algorithm.
Nevertheless, our argument that LH-MIN$^*$ is BQP-complete implies
that even its ``gapped'' version is unlikely to have an efficient
classical algorithm: the Hamiltonian
\begin{equation}
H' = H_{in} + H_{prop} + H_{clock}
\end{equation}
has a spectral gap $\delta' = \Omega(1/poly(n))$ above
its {\em unique} (non-degenerate) ground state $\ket{\eta}$
\cite{KSV,AvKLLR}, so we can choose the perturbation $H_{out}$ both
(a) large enough so that the promise gap $\Delta$ for $\lambda_1$ is
$\Omega(1/poly(n))$ and (b) small enough so that the spectral gap
$\delta$ of $H$ is $\Omega(1/poly(n))$ just like that of
$H'$.\footnote{This choice guarantees for both YES {\em and} NO instances
that the spectral gap above the ground state is large, although we
ignore the latter property henceforth.}  At this point, we may
reparametrize $(a,b) \mapsto (a, a+\delta)$ to conclude that the
special case of LH-MIN$^*$ which for a YES instance has a unique
eigenvalue at most $a$ and every other eigenvalue at least $b$ (let us
call this problem UNIQUE-LH-MIN$^*$) remains BQP-complete:
\begin{proposition}
The promise problem UNIQUE-LH-MIN$^*$ is BQP-complete under Karp
reduction.
\end{proposition}

Now consider the problem obtained by modifying UNIQUE-LH-MIN$^*$ so
that for a YES instance, the classical state of large inner product
with a low-energy state $\ket{\psi}$ is an unknown classical state
$\ket{b_1 b_2 \cdots b_n}$.  We might guess that it is QCMA-complete,
and this is essentially true: although the Karp reduction given by Kitaev
\cite{KSV} does not produce a gapped $H$ if there are multiple
classical witnesses, we can force it to do so by composing it with a
randomized reduction of the sort used by Valiant and Vazirani \cite{VV}.

Interestingly, it is not known how to apply the Valiant-Vazirani
technique to $H$ if its eigenvectors (as an unordered set of
orthogonal axes) are unknown and highly non-classical \cite{AGIK}.
Perhaps it is not possible:  there is some theoretical
evidence that a large spectral gap implies that the ground state
exhibits little long-range entanglement \cite{HNO} and is therefore
approximable by a succinct (classical) representation such as a
``matrix product state.''  If this were true generally, then one could
not reduce LH-MIN to gapped instances without also showing QMA=QCMA.

\paragraph{Further directions.}
Beyond those we have already mentioned, there are several LH-MIN
restrictions known to be complete for various subclasses of QMA:  If
we restrict the local terms $H_e$ of $H$ to be classical (i.e.,
diagonal), we obtain an NP-complete problem generalizing MAX-$k$-SAT.
If we restrict each $H_e$ to be a
projection matrix and set $a = 0$, we obtain the ``quantum $k$-SAT''
problem complete for QMA$_1$ \cite{Bra}.  If these projection matrices
are required to have nonnegative entries, we obtain the ``stoquastic
$k$-SAT'' problem complete for MA \cite{BBT}.  Determining the degree
to which each of these promise problems can be relaxed or tightened
while retaining the same computational complexity merits further
investigation.

\bibliographystyle{plain}
\bibliography{remarks}

\begin{thebibliography}{10}

\bibitem{AK}
S.~Aaronson and G.~Kuperberg.
\newblock Quantum versus classical proofs and advice.
\newblock {\em Theory of Computing}, 3(7):129--157, 2007.

\bibitem{AL}
D.~Abrams and S.~Lloyd.
\newblock A quantum algorithm providing exponential speed increase for finding
  eigenvalues and eigenvectors.
\newblock {\em Phys. Rev. Lett.}, 83:5162--5165, 1999.

\bibitem{AAEL}
D.~Aharonov, I.~Arad, E.~Eban, and Z.~Landau.
\newblock Polynomial quantum algorithms for additive approximations of the
  {P}otts model and other points of the {T}utte plane.
\newblock quant-ph/0702008.

\bibitem{AGIK}
D.~Aharonov, D.~Gottesman, S.~Irani, and J.~Kempe.
\newblock The power of quantum systems on a line.
\newblock In {\em Proc. 48th IEEE FOCS}, pages 373--383, 2007.

\bibitem{AJL}
D.~Aharonov, V.~Jones, and Z.~Landau.
\newblock A polynomial quantum algorithm for approximating the {J}ones
  polynomial.
\newblock In {\em Proc. 38th ACM STOC}, pages 427--436, 2006.

\bibitem{AvKLLR}
D.~Aharonov, W.~van Dam, J.~Kempe, Z.~Landau, S.~Lloyd, and O.~Regev.
\newblock Adiabatic quantum computation is equivalent to standard quantum
  computation.
\newblock {\em SIAM Journal on Computing}, 37(1):166--194, 2007.

\bibitem{Bra}
S.~Bravyi.
\newblock Efficient algorithm for a quantum analogue of 2-{SAT}.
\newblock quant-ph/0602108.

\bibitem{BBT}
S.~Bravyi, A.~Bessen, and B.~Terhal.
\newblock Merlin-{A}rthur games and stoquastic complexity.
\newblock quant-ph/0611021.

\bibitem{HNO}
H.~Haselgrove, M.~Nielsen, and T.~Osborne.
\newblock Entanglement, correlations, and the energy gap in many-body quantum
  systems.
\newblock {\em Phys. Rev. A}, 69:032303, 2004.

\bibitem{JW}
D.~Jenzing and P.~Wocjan.
\newblock A simple {P}romise{BQP}-complete matrix problem.
\newblock {\em Theory of Computing}, 3(4):61--79, 2007.

\bibitem{KSV}
A.~Kitaev, A.~Shen, and M.~Vyalyi.
\newblock {\em Classical and quantum computation}.
\newblock AMS, 2002.

\bibitem{KL}
E.~Knill and R.~Laflamme.
\newblock Quantum computation and quadratically signed weight enumerators.
\newblock {\em Inform. Process. Lett.}, 79(4):173--179, 2001.

\bibitem{VV}
L.~Valiant and V.~Vazirani.
\newblock N{P} is as easy as detecting unique solutions.
\newblock {\em Theoretical Computer Science}, 47:85--93, 1986.

\bibitem{WJB}
P.~Wocjan, D.~Janzing, and T.~Beth.
\newblock Two {QCMA}-complete problems.
\newblock {\em Quantum Information \& Computation}, 3(6):635--643, 2003.

\bibitem{WZ}
P.~Wocjan and S.~Zhang.
\newblock Several natural {BQP}-complete problems.
\newblock quant-ph/0606179.

\end{thebibliography}

\end{document}